# Mechanical behaviour of brain-skull interface (meninges) under shear loading through experiment and finite element modelling: Preliminary results

Sajjad Arzemanzadeh[1*][0000-0001-7381-1777], Karol Miller[1][0000-0002-6577-2082], Tim Rosenow[2][0000-0003-3727-5131], Sjoerd B. Vos[2][0000-0002-8502-4487], and Adam Wittek[1][0000-0001-9780-8361]

[1] Intelligent Systems for Medicine Laboratory (ISML), School of Engineering, The University of Western Australia, Perth 6009, WA, Australia
sajjad.arzemanzadeh@research.uwa.edu.au

[2] Western Australia National Imaging Facility, The University of Western Australia, Perth 6009, WA, Australia

**Abstract.** The brain-skull interface (meninges) plays a critical role in governing brain motion during head impacts, yet computational models often simplify this interface using idealized contact conditions due to limited experimental data. This study presents an improved protocol combining experimental testing and computational modelling to determine the mechanical properties of the brain-skull interface under shear loading. Brain tissue and brain-skull complex samples were extracted from sheep cadaver heads and subjected to shear loading. Magnetic resonance imaging (MRI) was used to obtain accurate 3D geometries of the samples, which were then used to create computational grids (meshes) for simulation of the experiments using finite element (FE) models to determine subject-specific properties of the brain tissue and brain-skull interface. A second-order Ogden hyperelastic model was used for the brain tissue, and a cohesive layer was employed to model the brain-skull interface. Our results indicate that a cohesive layer captures the force-displacement and damage initiation of the brain-skull interface. The calibrated cohesive properties showed consistent patterns across samples, with maximum normal tractions ranging from 2.8-3.4 kPa and maximum tangential tractions from 1.8-2.1 kPa. This framework provides a foundation for improving the biofidelity of computational head models used in injury prediction and neurosurgical planning by replacing arbitrary boundary conditions with formulations derived from experimental data on brain-skull interface (meninges) biomechanical behaviour.

**Keywords:** Brain-skull interface, Meninges, Brain tissue, Biomechanical properties under shear load, Finite element simulation.

## 1 Introduction

Finite element (FE) modelling of the human head has become a key technique in biomedical engineering and neuroscience enabling us to investigate the biomechanical responses of brain tissue to transient mechanical loads [1-3]. Creating computational biomechanics models of the head requires information about the head and brain geometry, boundary and loading conditions, and constitutive properties of the intracranial and extracranial components. Among these, accurate patient-specific geometry can be obtained from magnetic resonance images (MRIs), loading conditions can be specifically defined based on the target application, and brain material properties have been widely studied and refined over the years [4]. However, boundary conditions for the brain, which are determined by the biomechanical behaviour of the brain-skull interface (also referred to as meninges), remain underexplored. Ad-hoc boundary conditions, derived from parametric studies and supported only by limited and indirect experimental evidence, still represent the state-of-the-art [5, 6].

The brain-skull interface is a layered structure comprising dura mater, arachnoid, and pia mater (see **Fig. 1**) and provides both mechanical anchorage and physiological protection for the brain [7, 8]. The dura mater is the outermost layer, consisting of dense fibrous connective tissue which is attached firmly to the inner surface of the skull. The pia mater, the innermost layer, is tightly adherent and directly attached to the brain's outer surface. The subarachnoid space is a complex web-like structure sandwiched between the dura and pia mater. The exact anatomical structure and mechanical characteristics of these sublayers have long been an area of research as limited quantitative data are available [6, 7]. Sheep meninges are thinner, less complex, and their arachnoid villi and CSF drainage differ compared to human meninges [9-12]. However, the overall biological functions, anatomical structures, and biomechanical behaviour remain broadly similar to those of humans. Therefore, due to the ethical and technical constraints of conducting experiments on human meninges and brain tissue within a short time frame post-mortem, animal species with similar neuroanatomy, such as ovine, porcine and bovine, are often used [13-16].

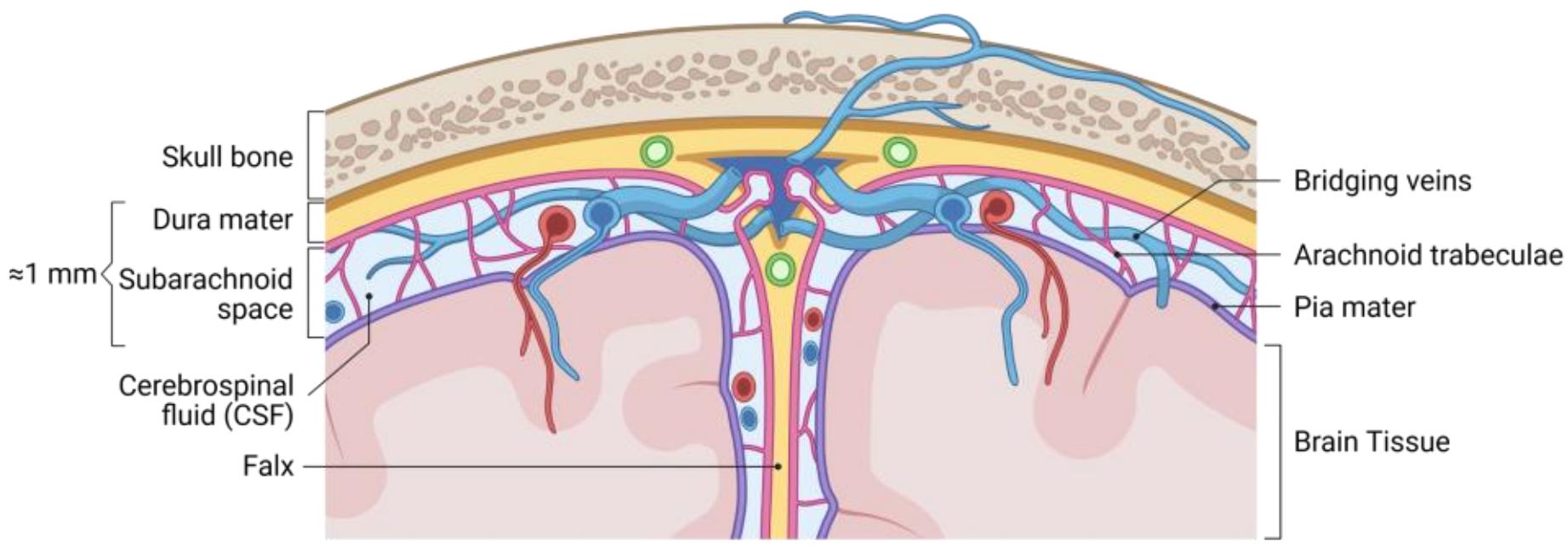

**Fig. 1.** Schematic of the brain-skull interface (meninges). For humans, the average thickness of the meninges is 1 mm approximately, other relative dimensions in the schematic are not intended to be scaled.

The thin and structurally complex brain-skull interface (**Fig. 1**) presents substantial challenges to direct biomechanical testing. Some studies have conducted tests on isolated meningeal layers, such as the dura or pia-arachnoid complex. Walsh et al. [17] investigated mechanical properties of the human dura mater under biaxial tensile loading and revealed that dura mater is heterogeneous with higher elastic moduli for both low and high strain regions in the medial-lateral direction compared to the anterior-posterior direction. Jin et al. [15] extracted pia-arachnoid samples from cadaver bovine heads and conducted in-plane tension, normal traction, and shear tests. They reported the elastic modulus of pia-arachnoid complex between 6-40 MPa under in-plane tension loading, 61-148 kPa under traction loading, and 11-22 kPa under shear loading, for strain rates between 0.05 and 100 $s^{-1}$.

Few attempts have been made to test the brain-skull interface as a whole unit rather than as isolated meningeal layers. MacManus et al. [18] removed the skull from porcine cadaver heads, tested the exposed meninges under dynamic indentation, and used FE analysis to calibrate viscoelastic properties of meninges. Here, we employ an even more cautious method that minimises the risk of damage to meninges when preparing the samples for experiments. In our approach, the dissection used to prepare the samples preserves the skull bone, brain-skull interface, and brain tissue within the sample (**Fig. 2**). It was preliminarily verified by Arzemanzadeh et al. [19] who conducted compressive and tensile experiments on brain tissue and brain-skull complex samples extracted from sheep cadaver heads. They used FE models to calibrate the subject-specific brain tissue mechanical properties and then investigated the brain-skull interface of the samples extracted from the same sheep cadaver head. They found that assuming a rigid connection or frictionless sliding contact between the brain and skull bone does not accurately represent the mechanical behaviour of the brain-skull interface.

In this work, we modify the experimental and modelling framework introduced in Arzemanzadeh et al. [19] and apply it to determine the mechanical behaviour of the brain-skull interface in shear. We present a new approach for implementing this behaviour in FE head models using a cohesive zone model (CZM). The 3D geometry of the samples was obtained from MRIs and used to construct geometrically accurate FE meshes for computational biomechanics models. These models were then used to analyse experimental results to determine the subject-specific constitutive properties of the brain tissue and mechanical behaviour of the brain-skull interface.

## 2 Methods

Skinned sheep cadaver heads were collected from an abattoir (Dardanup Butchering Company) and transported to the laboratory facilities at the Harry Perkins Institute of Medical Research, where sample preparation, MRI acquisition, and biomechanical testing were conducted, with The University of Western Australia Biosafety Approval RA/3/ F69199 and Harry Perkins Institute of Medical Research Institutional Biosafety Committee Approval RA 23/2023.

### 2.1 Sample Preparation

Two sets of cuboidal samples were extracted from each head: (i) brain tissue alone, and (ii) a brain-skull complex comprising brain tissue, brain-skull interface, and skull bone. All experiments were conducted within 24 hours of the animals' death, a timeframe shorter than postmortem delays known to significantly alter the mechanical characteristics of the brain tissue [20].

Cutting tools and tissue dissection procedure follow our previous work [19], but the region from which samples were extracted has been revised. In our previous study, samples were extracted from the middle of the head containing both hemispheres and therefore, bridging veins and falx had a dominant effect on the mechanical behaviour of the brain-skull interface. As our goal is to determine the mechanical properties of the meninges alone, we extracted samples from a single hemisphere so that the bridging veins and falx do not govern the overall behaviour of the interface. **Fig. 2a** shows the sheep cadaver skinned head before sample extraction with indicated cutting lines and **Fig. 2b** shows a schematic of the region from which samples were extracted with approximate dimensions of the samples. Part A was removed first to provide access to the area beneath part 1 and 2, which was then horizontally dissected in the transverse plane. Part 1 was used as the brain-skull complex sample and part 2 served as the brain tissue sample after the removal of the skull bone.

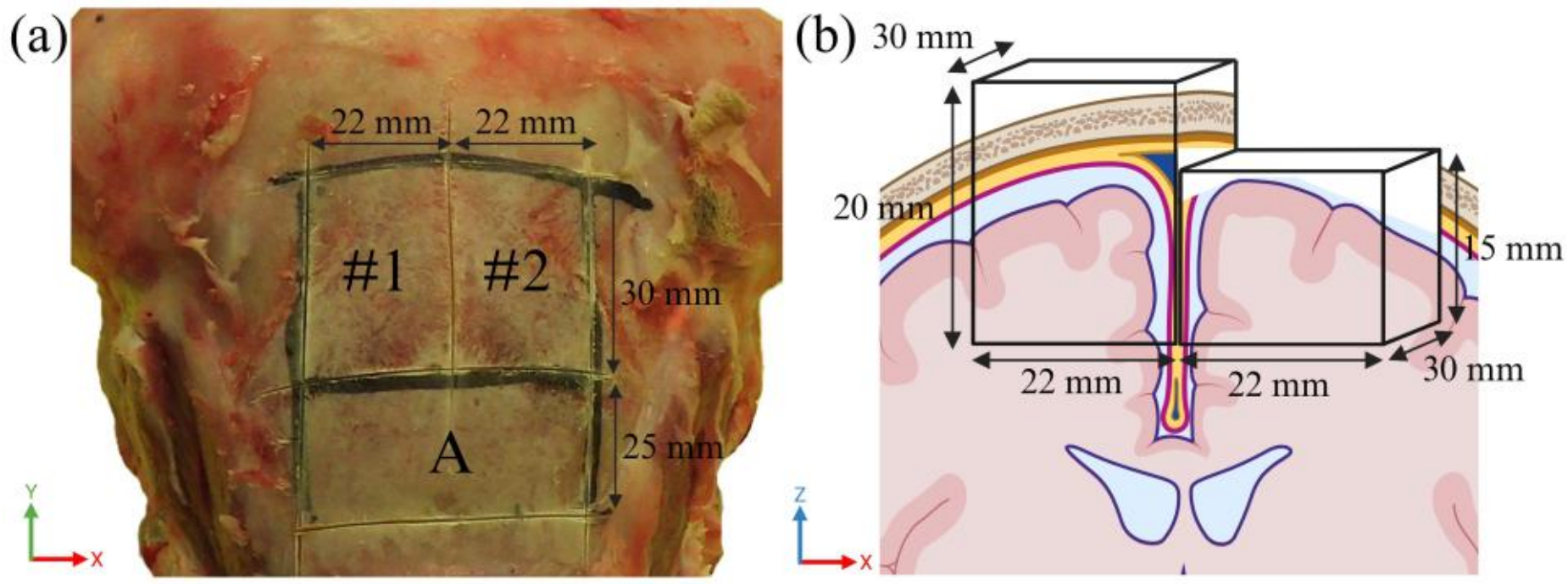

**Fig. 2.** (a) Sheep cadaver head before sample extraction with indicated cutting lines. 1: location of extraction of the brain-skull complex sample. 2: location of extraction of the brain tissue sample (with removal of the skull bone following the extraction). A: location where the brain tissue and skull bone were removed to provide access for horizontal cut and extraction of the brain-skull complex and brain tissue samples. (b) schematic of a cross-section of the head with approximate region and dimensions of the extracted samples. Left-hand-side: brain-skull complex sample. Right-hand-side: brain tissue sample.

After dissection of the brain-skull complex sample, an epoxy putty (Selleys' Knead It Aqua) was used to level the uneven skull surface and make the brain-skull interface parallel to the direction of loading. The dissection procedure has been designed to extract cuboidal samples with dimensions of 22 mm x 30 mm x 15 mm, for the brain tissue, and 22 mm x 30 mm x 20 mm for the brain-skull complex (**Fig. 2**). However, due to adhesion and very low stiffness of the brain tissue, there is significant risk of inducing damage during the tissue handling and cutting/dissection. We prioritised minimising this risk rather than achieving the cuboidal geometry. Therefore, MRIs were acquired to determine the actual 3D geometry of the brain tissue and brain-skull complex samples.

### 2.2 Acquisition of Sample Images

MRIs of the samples were acquired using a Siemens Healthineers Vida 3T MRI scanner at the Western Australia National Imaging Facility (NIF) node located at the Harry Perkins Institute of Medical Research (Nedlands, WA, Australia). We used a two-dimensional T2-weighted turbo spin echo sequence with repetition time = 15830ms, echo time = 93ms, averages = 11, slices = 64, turbo factor = 18, slice thickness = 0.6mm, matrix size = 320x220, and in-plane resolution = 0.6 mm x 0.6 mm.

### 2.3 Experiments

We conducted 6 experiments (three using brain tissue samples and three using the brain-skull complex samples) extracted from three sheep cadaver heads as indicated in **Table 1**, using the equipment and facilities available at the T3mPLATE Laboratory at Harry Perkins Institute of Medical Research (Nedlands, WA, Australia). The experimental setup is shown in **Fig. 3**. The UniVert (CellScale, www.cellscale.com) specialised portable biomaterial testing system driven by a stepper motor under closed loop control was used for shear tests. The testing system was equipped with a 2.5 N load-cell (miniature S-Beam Futek load-cell, accuracy of 0.1% of the maximum load which implies 2.5 mN) and displacement transducer. The force and loading head displacements were recorded with a sampling rate of 100 Hz. The loading and base platens were custom 3D-printed to fit the geometry of the samples and UniVert components. The experimental protocol follows the general guidelines of Miller [13, 21] and Miller and Chinzei [14, 22] for compressive and tensile testing, and has been modified for shear testing. The dorsal (cortical) surface of the brain tissue samples and outer skull surface for the brain-skull complex samples were glued to the base platen using a fast-curing cyanoacrylate glue (Selleys' super glue), a type of adhesive previously used in biomechanical testing of soft tissues [14, 21, 23]. The loading was applied to the ventral surface of the samples. To impose no-slip boundary conditions, the ventral surface of the sample was glued to the loading head using the same fast-curing cyanoacrylate glue. Cyanoacrylate glue cures quickly and forms a strong bond to both tissue and platen surfaces without penetrating them. Furthermore, being orders of magnitude stiffer than brain tissue, it does not deform under shear load and thus provides a nearly rigid connection. An adjustable height platform was used as the base of the test setup. To ensure adhesion, the base platen was moved 1 mm upwards after the initial contact of the ventral surface of the samples with the loading platen. It was held in this position for 30 s, then returned to the original contact point and held there for an additional 180 s to provide stress relaxation before the start of the shear tests. The time of 180 s was selected as it has been reported in the experimental literature as sufficient for brain tissue relaxation [24]. The loading was applied along the transverse plane which is along the positive direction of X axis for brain-skull complex samples and negative direction of X axis for brain tissue samples based on coordinate system shown in **Fig. 2**. The loading speed was set to 0.3 mm/s which implies a shear strain rate of approximately 0.02/s. All tests were conducted at room temperature (20˚C-22˚C).

**Table 1.** List of samples and their type (brain tissue and brain-skull complex). Each sheep cadaver head provided one brain tissue and one brain-skull complex sample.

| Sample | Type of sample | Sheep cadaver head |
|---|---|---|
| B1 | Brain tissue | Head1 |
| S1 | Brain-skull complex | Head1 |
| B2 | Brain tissue | Head2 |
| S2 | Brain-skull complex | Head2 |
| B3 | Brain tissue | Head3 |
| S3 | Brain-skull complex | Head3 |

The experiments were recorded using three digital cameras: (1) High resolution DFK 33UX264 scientific camera (resolution of 2048 x 2048, 15 Frames/s) that provided the frontal view of the test samples; (2) Canon LEGRIA HF G70 camcorder (resolution of 3840x2160, 25 Frames/s) on the right side of the setup and (3) Sony HDR-CX625 camcorder (resolution of 1920x1080, 25 Frames/s) on the left side of the setup that provided the general oblique views of the test samples. These recordings were used to investigate the deformation of brain tissue and behaviour of brain-skull interface under shear loading.

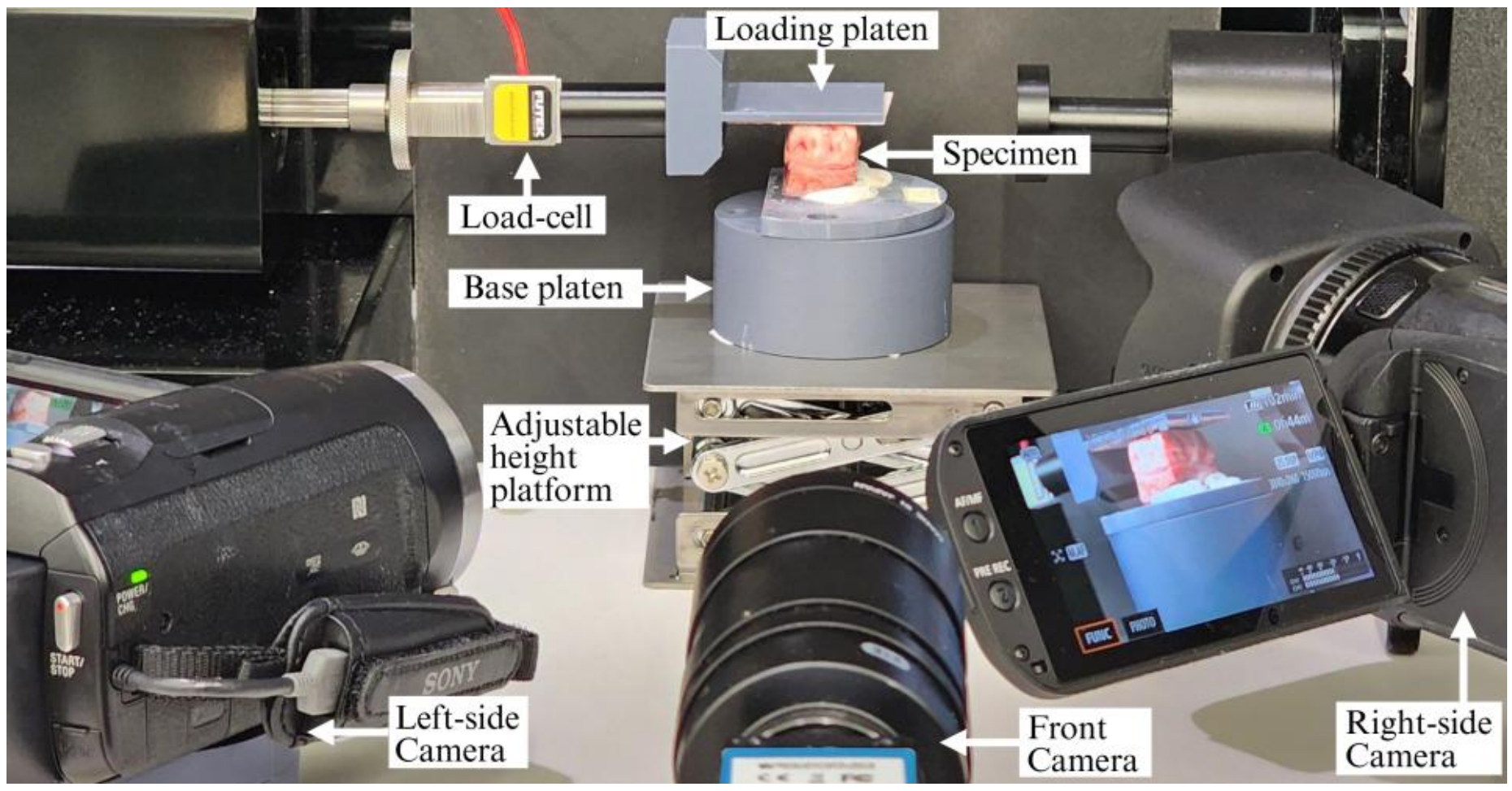

**Fig. 3.** Experimental setup for shear test on brain tissue and brain-skull complex samples.

### 2.4 Finite Element Models for Determining Material Parameters of the Brain Tissue and Brain-skull Interface

The 3D Slicer software platform for image computing and 3D visualisation [25] was used to semi-automatically segment the MRIs of the samples to extract their accurate 3D geometry. Finite element meshes of the samples were then created using these geometries. Meshing was performed using Coreform Cubit 2025.1 (Coreform,

www.coreform.com/products/coreform-cubit), a finite element mesh generator. The meshes were then imported into Abaqus 2023 finite element software (Dassault Systèmes, www.3ds.com/products/simulia/abaqus). An element size of 0.5 mm was used. This allowed us to generate high-quality, fully hexahedral meshes that accurately represent 3D sample geometry determined from the MRIs. The average scaled Jacobian was above 0.95 and minimum scaled Jacobian was 0.28 for the brain tissue elements in all samples (**Fig. 4**) [26]. All simulations were conducted using the Abaqus explicit dynamics non-linear finite element solver with automated time stepping. To prevent volumetric locking, we used under-integrated hexahedral element formulation (C3D8R element type in Abaqus) with stiffness-based hourglass control for the brain tissue and skull, and hexahedral cohesive element formulation (COH3D8 element type in Abaqus) for brain-skull interface (SIMULIA User Assistance 2023/Abaqus manual).

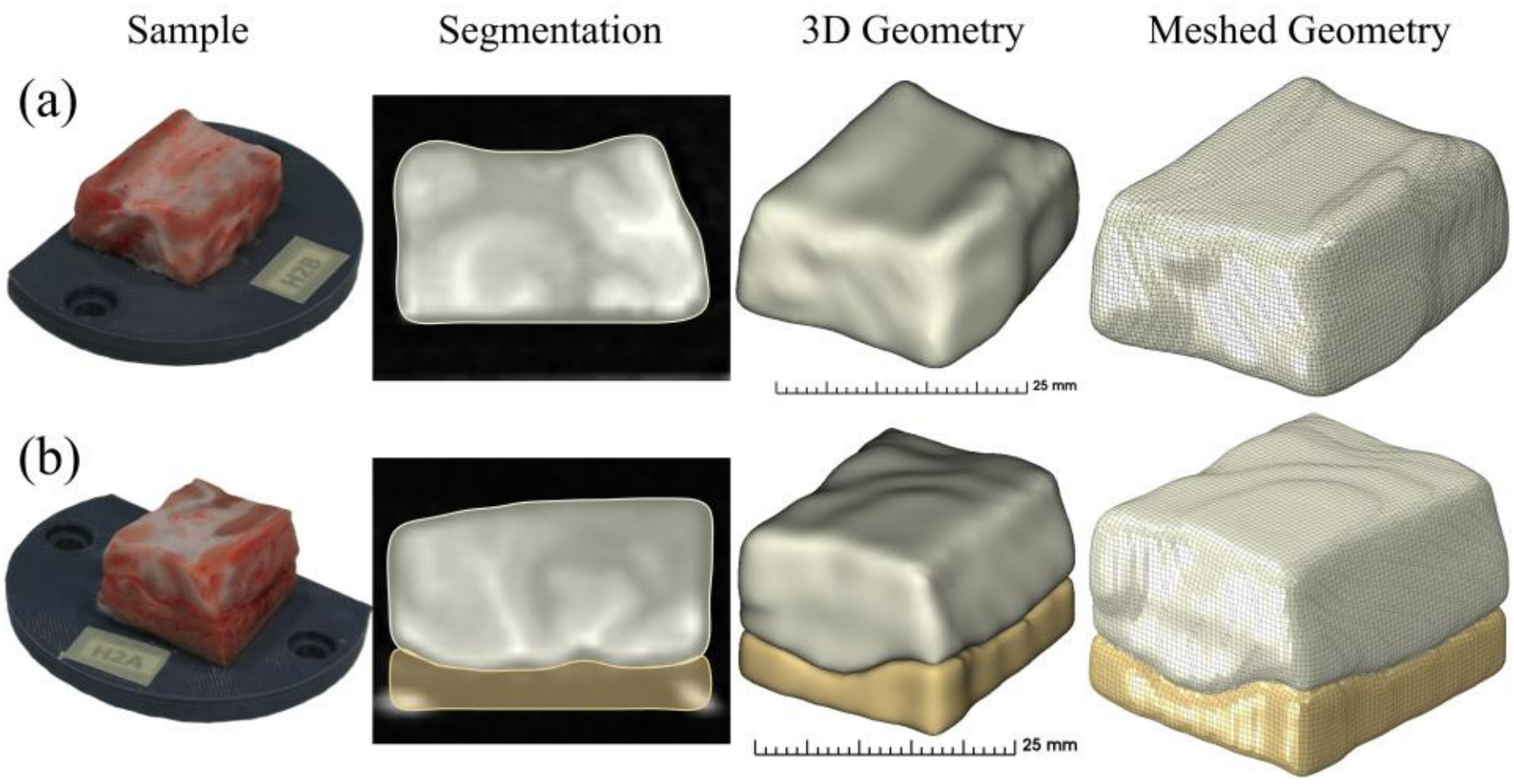

**Fig. 4.** (a) Brain tissue sample (B1) and (b) brain-skull complex sample (S1) used in shear test. First column: experimental samples, second column: the MRI of the samples with highlighted material segmentation, third column: 3D representation of the segmentation, and fourth column: meshed geometries imported into Abaqus. White colour: brain tissue; Yellow colour: skull bone.

In the FE models of brain tissue samples, the bottom nodes of the samples were rigidly constrained as in the experiments the bottom (dorsal) surfaces of the samples were glued to the base platen. In the FE models of brain-skull complex samples, the skull bones were rigidly constrained as in the experiments the outer surface of the skull bones was glued to the base platen. The top surface of all samples was bonded to the loading head using fast-curing cyanoacrylate glue to constrain movement to the horizontal direction. Therefore, in all FE models, loading was defined by prescribing the displacement (rate of 0.3 mm/s) in the horizontal direction on the model nodes representing the top surfaces of the samples. The displacement was applied using a smooth step procedure that uses a $5^{th}$ order polynomial in the Abaqus finite element software (SIMULIA User Assistance 2023/Abaqus manual).

### 2.4.1 Constitutive Models and Parameters

The skull bone was modelled as rigid since it is orders of magnitude stiffer than the brain tissue. A second-order Ogden-type hyperelastic model ($N$=2) [27] has been used to describe the material behaviour of the brain tissue:

$$W(\overline{\lambda_1}, \overline{\lambda_2}, \overline{\lambda_3}) = \sum_{i=1}^{N} \frac{2\mu_i}{\alpha_i^2} \left( \overline{\lambda_1}^{\alpha_i} + \overline{\lambda_2}^{\alpha_i} + \overline{\lambda_3}^{\alpha_i} - 3 \right) + \sum_{i=1}^{N} \frac{1}{D_i} (J_{el} - 1)^{2i}, \quad (1)$$

Where $\overline{\lambda_i}$ are deviatoric principal stretches, $J_{el}$ is the elastic volume change, $\mu_i$ are the shear moduli, $\alpha_i$ are the dimensionless constants, and $D_i$ are material constants determining compressibility. The initial shear modulus is given by

$$\mu_0 = \sum_{i=1}^{N} \mu_i, \quad (2)$$

and the initial bulk modulus is given by

$$K_0 = \frac{2}{D_1}. \quad (3)$$

As the brain is nearly incompressible, a Poisson's ratio of 0.49 was used [4]. A unique set of material constants ($\mu_i$ and $\alpha_i$) was determined for shear strain of up to 0.3 shear strain for brain tissue of each sheep cadaver head using optimisation. We applied sequential least squares programming (SLSQP) algorithm [28] to minimise the sum of absolute difference between force-displacement curves observed in the experiments and obtained from the FE models. Then, the subject-specific material constants of the brain tissue were used as brain tissue material properties to model the experiments on the brain-skull complex samples extracted from the same sheep cadaver head to determine the material properties of the brain-skull interface.

To model the behaviour of the brain-skull interface, we used a cohesive zone model (CZM), an approach often used when modelling layers of material with anisotropic properties, such as adhesives [29, 30]. To model the damage initiation and propagation to failure of the brain-skull interface, we use a traction-separation-based CZM (SIMULIA User Assistance 2023/Abaqus manual). In the traction-separation-based CZM, the traction transmission capability across cohesive zone degrades as the nominal displacement increases. Traction in cohesive elements can be defined as:

$$\begin{Bmatrix} t_n \\ t_s \\ t_t \end{Bmatrix} = \begin{pmatrix} E_{nn} & 0 & 0 \\ 0 & E_{ss} & 0 \\ 0 & 0 & E_{tt} \end{pmatrix} \begin{Bmatrix} \varepsilon_n \\ \varepsilon_s \\ \varepsilon_t \end{Bmatrix}, \quad (4)$$

where $E_{nn}$, $E_{ss}$, and $E_{tt}$ are linear elastic traction moduli, and $t_n$, $t_s$, and $t_t$ are the nominal tractions in the normal and the two local shear directions, respectively; while the quantities $\varepsilon_n$, $\varepsilon_s$, and $\varepsilon_t$ represent the corresponding nominal strains. Maximum nominal stress ratio was chosen as the damage initiation criterion, expressed as:

$$\max \left\{ \frac{\langle t_n \rangle}{t_n^0}, \frac{t_s}{t_s^0}, \frac{t_t}{t_t^0} \right\} = 1. \quad (5)$$

To model the degradation of interface stiffness following damage initiation, an energy-based damage evolution was employed, representing the energy dissipated during failure (G). We did not attempt to calibrate elastic moduli of the brain-skull interface as we only conducted three experiments on brain-skull complex samples under shear loading. Therefore, the elastic moduli of the brain-skull interface were obtained from the literature for the pia-arachnoid complex under comparable strain rates. Accordingly, based on the findings of Jin et al. [15], an elastic traction modulus of 61 kPa was assigned to $E_{nn}$ and a shear modulus of 11 kPa was used for both $E_{ss}$ and $E_{tt}$ in our FE models of brain-skull complex samples. The shear properties of the brain-skull interface were assumed to be isotropic in the tangential directions. Consequently, the nominal tractions ($t_n$ and $t_s$= $t_t$, as defined in Equation 4) and fracture energy ($G$) were the cohesive properties of the brain-skull interface calibrated to match experimental force-displacement results of the brain-skull complex samples under shear loading.

## 3 Results

For the brain tissue samples subjected to shear loading in this study, the force-displacement characteristics obtained from the FE models of the samples were close to the characteristics observed in the experiments (**Fig. 5**). The maximum force difference between the computed and experimentally measured forces was 4.8% for sample B1, 6.6% for sample B2, and 7.8% for sample B3. These results confirm the validity of our selection of the second-order hyperelastic Ogden material model (see Equation 1) for the brain tissue and accuracy of our procedures for determining the parameters for this model described in section 2.4.

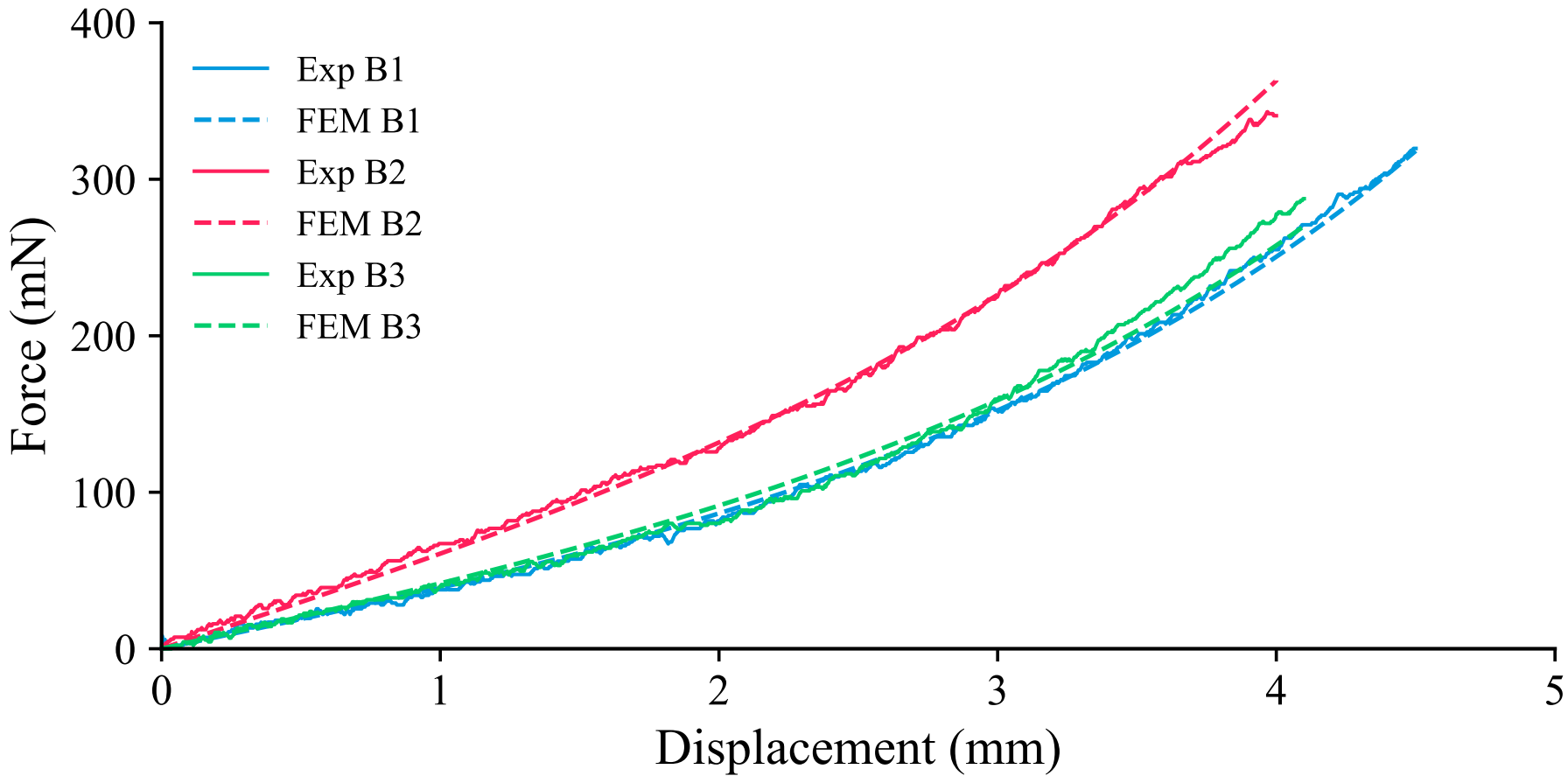

**Fig. 5.** Force-displacement characteristics obtained from experiments and FE models of shear tests conducted on brain tissue samples. B1, B2, and B3 stand for brain tissue samples extracted from first, second, and third sheep cadaver heads, respectively.

The initial shear modulus ($\mu_0$) and first dimensionless parameter ($\alpha_1$) of the Ogden model, that dominate the force-displacement characteristics at low strains, which we determined here (**Table 2**) are close to $\mu_0$=1200 Pa and $\alpha_1 = -6.3$ we previously reported for the sheep brain tissue under tensile and compressive loading [19]. For the three brain tissue samples analysed here, the ratio of initial shear modulus standard deviation to the average initial shear modulus was 17% which is less than the variations reported in the literature [31].

**Table 2.** Parameters of the second-order Ogden hyperelastic material model (see Equation 1) of the brain tissue under shear loading. The parameters were determined for shear strain up to 0.3. B1 indicates the brain tissue sample extracted from the first sheep cadaver head, B2 - the sample from the second head, and B3 - the sample from the third head.

| Sample | $\mu_1$ (Pa) | $\alpha_1$ | $\mu_2$ (Pa) | $\alpha_2$ | $\mu_0$ (Pa) |
|---|---|---|---|---|---|
| B1 | 800.0 | -8 | 386.7 | 16 | 1186.7 |
| B2 | 1210.8 | -8 | 466.4 | 16 | 1677.2 |
| B3 | 821.6 | -8 | 599.9 | 16 | 1421.5 |

For the brain-skull complex samples, the force initially increases with shear displacement, then it plateaus and starts to gradually decrease (**Fig. 6**). Analysis of the camera recordings of the experiments conducted in this study clearly indicates that this force decrease is associated with the mechanical failure of the brain-skull interface within the subarachnoid space leading to separation of the brain from the skull at a shear strain of 0.3 to 0.4 (see **Fig. 6** and **Fig. 7**). As explained in Section 2.4.1, we represented this behaviour of the brain-skull interface using a layer of cohesive elements implemented in Abaqus finite element code. The cohesive zone model parameters that describe the failure behaviour of the brain-skull interface (maximum normal traction $t_n$, maximum tangential tractions $t_s$ and $t_t$, and fracture energy $G$, Equations 4 and 5) determined in this study exhibit the ratio of standard deviation to mean below 20% (**Table 3**). This is consistent with the variation in the maximum shear force and shear displacement at which the force maximum occurs (and force decrease starts) observed in our experiments using the brain-skull complex samples (**Fig. 6**).

**Fig. 7** shows the snapshots of experiment and FE model of the shear test on S1 sample. General behaviour of the force-displacement characteristics of the brain-skull complex samples obtained from our FE models using cohesive elements was consistent with the experimental results up to the shear displacement of between 5 and 6 mm, at which point the interface failure was evident in the experiments. This is consistent with good qualitative agreement between the general deformation and kinematics of the brain-skull complex samples predicted using our finite element models and recorded using high resolution cameras (see **Fig. 3** for the camera recording setup) during the experiments (**Fig. 6** and **Fig. 7**).

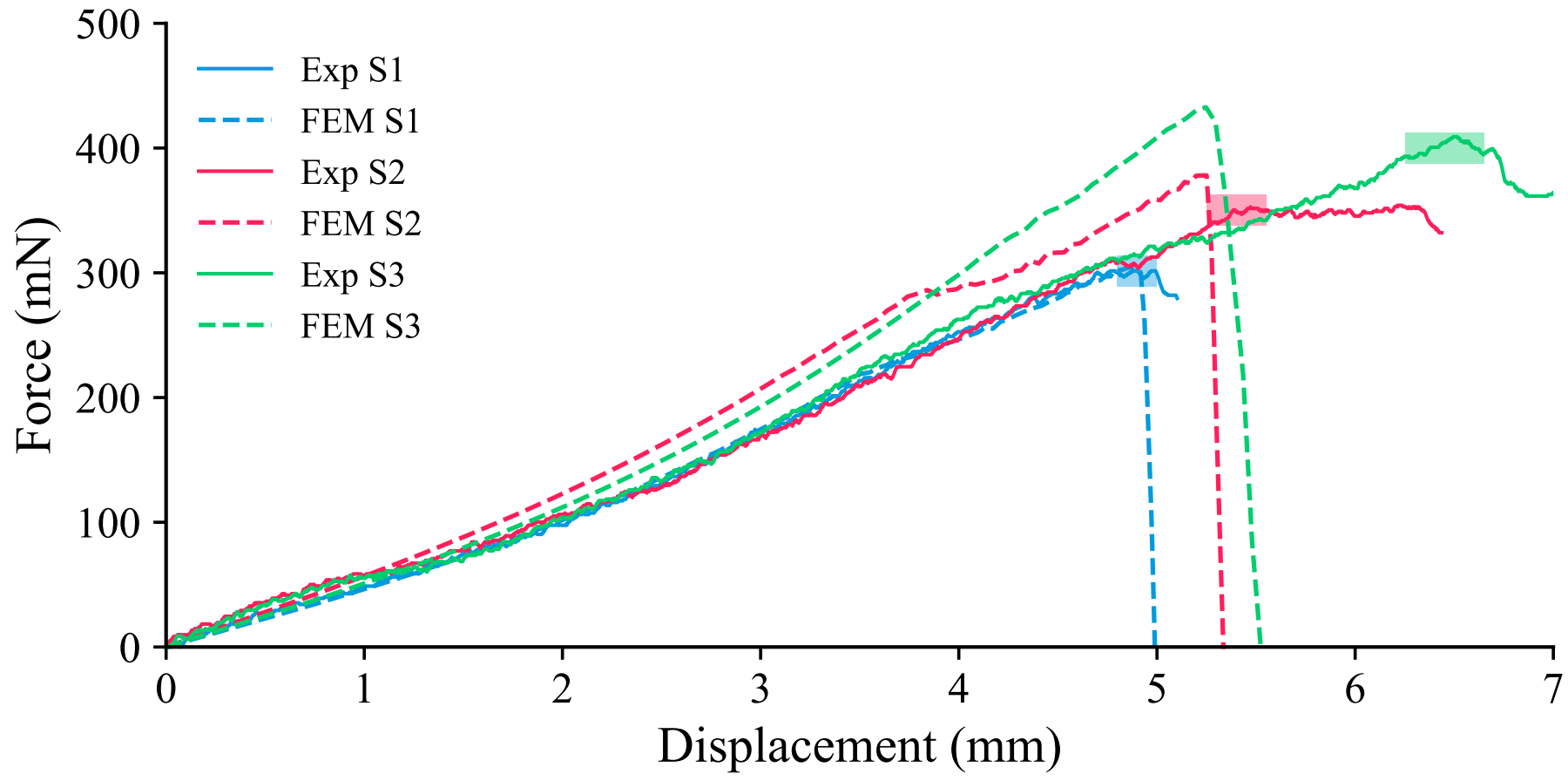

**Fig. 6.** Force-displacement characteristics obtained from experiments and FE models of shear tests. S1, S2, and S3 stand for brain-skull complex samples extracted from three sheep cadaver heads. Failure of the brain–skull interface in experiments has been detected through inspection of front- and side-view cameras recordings and highlighted with rectangles in the same colour as the samples.

**Table 3.** Parameters of cohesive zone (see section 2.4.1) for modelling the brain-skull interface determined from the experiments using the brain-skull complex samples under shear loading conducted in this study.

| Sample | Maximum normal traction (kPa) | Maximum tangential traction (kPa) | Fracture energy (N.m$^{-1}$) | Failure force difference percentage (%) |
|---|---|---|---|---|
| S1 | 3.0 | 2.1 | 0.48 | 1.0 |
| S2 | 3.4 | 1.9 | 0.54 | 6.7 |
| S3 | 2.8 | 1.8 | 0.7 | 5.7 |

For shear displacement of up to 1.0-1.5 mm, the force-displacement characteristics obtained from our models very closely matched those recorded in the experiments (**Fig. 6**), and the difference between the maximum force computed using the models and measured in the experiments was within 2.5%-10% range (**Table 3**). However, in all our models, rapid force-drop to zero occurred following the brain-skull interface failure, whereas a gradual force decrease was observed in the experiments (**Fig. 6**). It is unlikely that this discrepancy between modelling and experimental results was solely caused by the limitations when determining cohesive zone model parameters (maximum normal traction $t_n$, maximum tangential tractions $t_s$ and $t_t$, and fracture energy $G$, Equations 4 and 5) that control the onset and evolution of failure of the brain-skull interface using the experiments conducted in this study. It is likely to be related to challenges of modelling complex interactions between the brain and skull following the brain-skull interface failure. Visual observations made during the current experiments and sample inspection after the experiments strongly suggest that the interface failure was initiated within the subarachnoidal space as, after the interface failure, the pia mater was present on the outer surface of the brain tissue and the dura mater remained intact and firmly attached to the skull. Therefore, adhesion between the pia and dura tissues was likely to generate shear/tangential (along the skull bone surface) force when the brain outer surface moved/"slide" along the skull bone following the interface failure. Such adhesion was not included in our models of the brain-skull complex. Consequently, the tangential forces due to the interactions between the brain and skull following the failure of cohesive elements representing the brain-skull interface were close to zero.

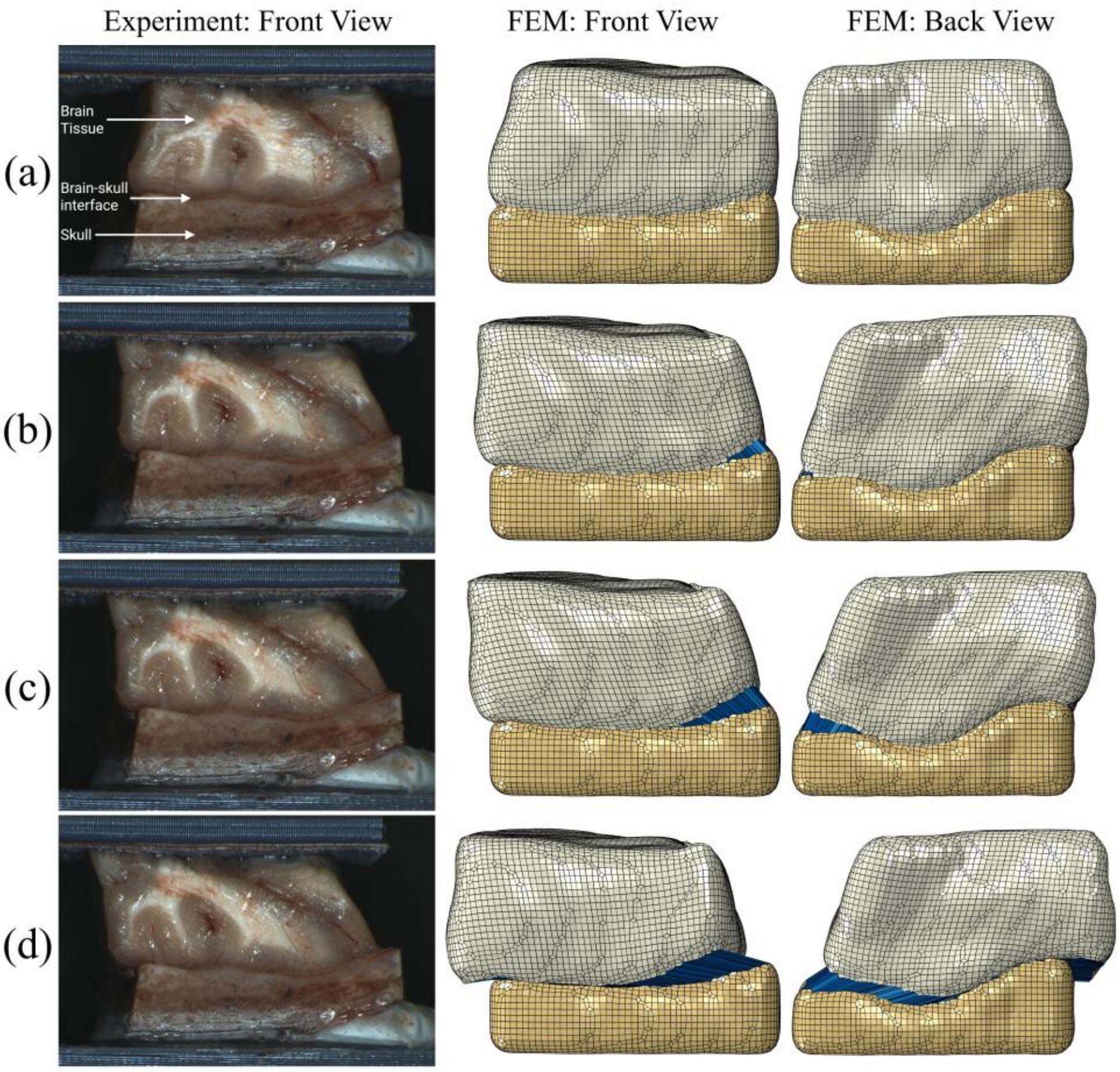

**Fig. 7.** Brain-skull complex sample (S1) under shear loading: (a) at the beginning of the test, (b) 3 mm tangential displacement, (c) beginning of the failure of the brain-skull complex sample (3.8 mm tangential displacement), and (d) after complete failure of the brain-skull interface (5.1 mm tangential displacement). First column is the front view of the experiment, second column is the front view of the FE model, and third column is the back view of the FE model.

## 4 Discussion

This study is one of very few attempts to determine the biomechanical behaviour of the brain-skull interface (meninges) that govern the interactions between the brain and skull, and boundary conditions for the brain. It focuses on the behaviour under shear loading and uses the general framework introduced in our previous studies of the biomechanical behaviour of the brain-skull interface under compression and tension [19, 32]. There are two important differences between this framework and approaches typically used in the field. First, we conducted our experiments on intact brain-skull interface. Whereas the approaches described in the literature use separate meningeal layers extracted from the brain-skull interface, which increases the risk of inducing accidental damage when dissecting the delicate meninges [15]. Second, in contrast to the vast majority of the studies conducted by others, that assume regular sample geometry [13, 14, 21, 22, 31-33], we used the actual sample geometry determined from magnetic resonance images (MRIs) when analysing our experiments. Although we did not attempt to quantify here the effects of the assumptions regarding the sample geometry on determining the brain tissue and brain-skull interface properties, a qualitative appreciation of possible drawbacks resulting from such assumptions can be gained from comparison of the sample photographs (first column in **Fig. 4**) and sample geometry extracted from the MRIs (third and fourth columns in **Fig. 4**). While the sample appears fairly regular in the photographs, 3D geometry determined from the MRIs through image segmentation reveals substantial departure from the nominal cuboidal geometry.

We conducted six experiments using three brain-skull complex samples (consisting of the brain tissue, brain-skull interface, and skull bone) and three brain tissue samples extracted from three sheep cadaver heads. One brain tissue and one brain-skull complex sample were extracted from each sheep cadaver head and subjected to shear loading. First, subject-specific material constants of the brain tissue were obtained through comprehensive biomechanical modelling of the experiments using non-linear explicit dynamics finite element (FE) analysis. Subject-specific material constants of the brain tissue were subsequently used in comprehensive modelling of the experiments using the brain-skull complex samples extracted from the same head. Observations made during the present experiments and our previous study in which the brain-skull complex tissue samples were subjected to tension [19] suggest that the brain-skull interface failure occurs within the subarachnoidal space while pia and dura remain largely intact and attached to the brain outer surface and skull respectively. Adhesion between pia and dura continues to generate force when the brain outer surface slides on the dura under shear loading following the interface failure, which is a phenomenon not included in the cohesive layer model we used here for the brain-skull interface. Therefore, incorporating this contact could be a possible improvement to better capture the gradual post-failure force decrease.

It should be noted that as the overall anatomical structure and biomechanical behaviour of brain tissue and meninges in sheep are broadly analogous to those of humans, one expects similar trends but different absolute values for humans [8].

In conclusion, the results of this study appear to confirm validity and feasibility of our experimental and modelling framework for discovering and quantitatively describing the biomechanical behaviour of the brain-skull interface (meninges) with the interface modelled using a layer of cohesive elements.

However, caution is needed as the parameters of cohesive layer were obtained from the results of only six experiments in which the brain tissue and brain-skull complex samples (consisting of the brain tissue, brain-skull interface, and skull bone) were subjected to shear loading. Providing more comprehensive answers regarding the brain-skull biomechanical behaviour, methods for modelling this behaviour, and specific model parameters would require experiments on a larger number of samples conducted under three loading conditions: tension, compression, and shear.

**Acknowledgements.** This research was supported by the Australian Government through the Australian Research Council's Discovery Projects funding scheme (Project DP230100949). Author S. A. acknowledges support of the University of Western Australia Scholarship for International Research Fees and University of Western Australia (UWA) - Intelligent Systems for Medicine Laboratory (ISML) Higher Degree by Research Scholarship in Computational Biomechanics (partly funded from Australian Research Council's ARC Discovery Project DP230100949). The authors acknowledge the facilities and scientific and technical assistance of the Western Australia National Imaging Facility (WA NIF) and thank Dr Tim Rosenow and Dr Sjoerd B. Vos of WA NIF and The University of Western Australia (UWA) for their support with image acquisition. The authors also thank Associate Professor Elena Juan Pardo, Head of the Translational 3D Printing Laboratory for Advanced Tissue Engineering (T3mPLATE) at the Harry Perkins Institute of Medical Research (Nedlands, WA, Australia), and the T3mPLATE laboratory personnel for providing access to the experimental equipment. The authors acknowledge Associate Professor Stuart Hodgetts (Spinal Cord Repair Laboratory, School of Human Sciences, The University of Western Australia) for his advice on sample dissection and storage, Ms Ulani Hayter Otaola, a student at the School of Engineering, The University of Western Australia, for her assistance with conducting the experiments, and Mr Alan Grant of Dardanup Butchering Company (Bunbury, Western Australia, Australia) for help in obtaining sheep cadaver heads used in this study.